\documentclass[prb,twocolumn,superscriptaddress,showpacs]{revtex4-1}
\usepackage{graphicx}
\usepackage{siunitx}
\usepackage{bm}
\usepackage{color}
\usepackage{amsmath}
\usepackage{amssymb}
\usepackage{multirow}

\newcommand{\DMMnF}{(CH$_3$)$_2$NH$_2$[Mn(HCO$_2$)$_3$]}
\newcommand{\DMMnFd}{(CD$_3$)$_2$ND$_2$[Mn(DCO$_2$)$_3$]}

\begin{document}

\title{Magnetic structure and spin wave excitations in the multiferroic magnetic metal-organic framework \DMMnFd}

\author{H. C. Walker}
\email{helen.c.walker@stfc.ac.uk}
\affiliation{ISIS Neutron and Muon Source, Rutherford Appleton Laboratory, Chilton, Didcot OX11 0QX, United Kingdom}
\author{H. D. Duncan}
\altaffiliation[Current address: ]{Centre for Science at Extreme Conditions, University of Edinburgh, Peter Guthrie Tait Rd, Edinburgh EH9 3FD, United Kingdom}
\affiliation{Centre for Condensed Matter and Materials Physics, School of Physics and Astronomy, Queen Mary University of London, Mile End Road, London E1 4NS, United Kingdom}
\author{M. D. Le}
\affiliation{ISIS Neutron and Muon Source, Rutherford Appleton Laboratory, Chilton, Didcot OX11 0QX, United Kingdom}
\author{D. A. Keen}
\affiliation{ISIS Neutron and Muon Source, Rutherford Appleton Laboratory, Chilton, Didcot OX11 0QX, United Kingdom}
\author{D. J. Voneshen}
\affiliation{ISIS Neutron and Muon Source, Rutherford Appleton Laboratory, Chilton, Didcot OX11 0QX, United Kingdom}
\author{A. E. Phillips}
\email{a.e.phillips@qmul.ac.uk}
\affiliation{Centre for Condensed Matter and Materials Physics, School of Physics and Astronomy, Queen Mary University of London, Mile End Road, London E1 4NS, United Kingdom}

\date{\today}

\begin{abstract}
  We report the magnetic diffraction pattern and spin wave excitations in \DMMnFd\ measured using elastic and inelastic neutron scattering. The magnetic structure is shown to be a G-type antiferromagnet with moments pointing along the $b$ axis. By comparison with simulations based on linear spin wave theory, we have developed a model for the magnetic interactions in this multiferroic metal-organic framework material. The interactions form a three-dimensional network with antiferromagnetic nearest-neighbour interactions along three directions of $J_1=-0.103(8)$~meV, $J_2=-0.032(8)$~meV and $J_3=-0.035(8)$~meV.
\end{abstract}

\maketitle

\section{Introduction}
For the past ten years there has been considerable interest in magnetoelectric multiferroics, both from the fundamental viewpoint and driven by the desire to create functional systems for modern technologies, such as multiferroic data storage \cite{Scott}. The search for optimal systems has revealed that traditional magnetic compounds are currently unable to provide the flexibility and complex functionality required. This has resulted in attention being turned to dense metal organic frameworks (MOFs), which present a versatile platform for the realization of complex magnets due to the high tailorability and tunability arising from their discrete molecular building-block nature \cite{Kurmoo}. However, the magnetic properties of such materials depend on their precise magnetic interactions, which are often poorly understood. In particular, while bulk magnetometry is routinely performed on newly reported materials, magnetic diffraction measurements are far less often reported \cite{lawler_probing_2015}, and magnetic spectroscopy is rarer still. If we are to develop magnetic MOFs for future applications it is vital that we rectify this situation, and obtain a greater understanding of the origin of their properties.

\begin{figure}[htp]
    \begin{center}
        \includegraphics[width=0.975\linewidth]{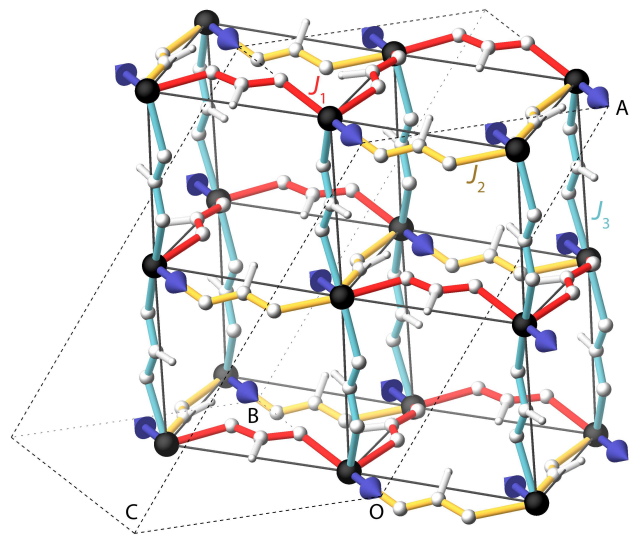}
    \end{center}
    \caption{\label{fig:struc} Simplified low temperature $Cc$ crystallographic structure of \DMMnFd\ showing the Mn ions linked together \emph{via} formate ion bridges, coloured red, yellow and cyan for the three nearest neighbour magnetic interactions. The network is composed of quasi-cubes with a G-type antiferromagnetic ordering of the Mn moments. (Dimethylammonium ions, which sit within these quasi-cubes, have been omitted for clarity.)}
\end{figure}

The dimethylammonium transition metal formates (DM$M$F) are one such family of magnetic MOFs displaying intriguing magnetic properties \cite{WangX,Jain08,Baker,WangZ,yadav_order-disorder_2016}. They have a three-dimensional perovskite structure, with the dimethylammonium sitting in the pore between the formate linked oxygen octahedra coordinating the metal ions (Figure~\ref{fig:struc}). Like their inorganic analogues, these materials may exhibit both electric and magnetic ordering, and their properties can be tuned by preparing solid solutions with substitution of either the metal \cite{maczka_effect_2015, maczka_structural_2016} or the organic cation \cite{kieslich_2016_tuneable}. Understanding the origins of these materials' electric and magnetic properties would allow this tunability to be exploited to design materials with targeted functionalities.

Transition metal ions can be linked \emph{via} a variety of different bridging modes by carboxylate ions, of which formate is the smallest. The metal-carboxylate bonds are conventionally described as \emph{syn} if the M--O--C--O dihedral angle is closer to zero, and \emph{anti} if this angle is near to $180^\circ$. Using this nomenclature, \emph{syn-anti} geometries tend to display ferromagnetic coupling, while \emph{syn-syn} and \emph{anti-anti} geometries are antiferromagnetic \cite{Colacio}. Depending on the carboxylate bridge, this can lead to 0, 1, 2 or 3 dimensional complexes\cite{WangX}. In DM$M$F, for $M=$Mn, Co, Fe and Ni, the transition metals are connected to the six nearest neighbours \emph{via} identically sized \emph{anti-anti} formate bridges. The nearest neighbour coupling is therefore predicted to be antiferromagnetic and they are reported to form canted antiferromagnets below $8.5$~K, $14.9$~K, $20$~K and $35.6$~K respectively, resulting in weak ferromagnetism. For $M=$Co and Ni there is an additional spin reorientation at $13.7$~K and $14.3$~K respectively\cite{WangX}.

DMMnF is the first reported perovskite metal-organic multiferroic \cite{Jain09}. It is a type-I multiferroic, since the origin of the magnetic and electric ordering differs: the former comes from the interactions of the Mn$^{2+}$ ions mediated by the formate linkers, while the latter comes from rotation of the (CH$_3$)$_2$NH$_2^+$ ions within the framework. A Curie-Weiss fit to the inverse magnetic susceptibility gives an effective paramagnetic moment of $5.94\mu_B$ consistent with that expected for $S=\tfrac52$ Mn$^{2+}$ and a negative Curie-Weiss constant of $-16.3$~K, confirming the presence of antiferromagnetic interactions \cite{WangX}. Based on molecular field theory, this temperature can be converted into an estimate for the exchange interaction of $-0.04$~meV. A small hysteresis loop is observed in the isothermal magnetisation, indicating that DMMnF is a weak ferromagnet, with an estimated spin canting angle of $0.08^\circ$\cite{WangX}. The ferroelectricity is driven by hydrogen bonds forming between the dimethylammonium protons and formate group oxygens at 183~K as they undergo a structural transition from $R\bar{3}c$ to $Cc$\cite{Sanchez}, when the dimethylammonium cation orders into one of three sites\cite{Duncan}. Since the ferroelectricity is unrelated to the magnetic order, in contrast to improper ferroelectric multiferroics like TbMnO$_3$\cite{Kimura,Cheong}, and given the disparity in the two ordering temperatures, any magnetoelectric coupling is likely to be weak, and its presence or absence has been the subject of some debate \cite{WangW,Abhyankar}. It appears that there is little to no magnetoelastic effect \cite{Abhyankar,Thomson}.

In this article we report the results of our powder elastic and inelastic neutron scattering study of perdeuterated dimethylammonium manganese formate. The results are compared with simulations for different exchange interaction models. 
Powder inelastic neutron scattering has proven to be a highly effective technique to probe magnetism in the inorganic perovskites \cite{mcqueeney_2008_determination,Holbein}, but has only rarely been used to study magnetism in formate frameworks\cite{Stride}. We show here that this technique is equally effective at revealing the behaviour of their metal-organic analogues.

\section{Experimental Details}
A $1.3$~g powder sample of perdeuterated DMMnF was synthesized according to the published solvothermal method\cite{WangX}, by heating a D$_2$O/CDON(CD)$_3$ solution of MnCl$_2\cdot$H$_2$O in a pressure vessel. Deuteration was necessary in order to avoid the large incoherent cross section of $^1$H nuclei.

Neutron diffraction measurements were performed using the GEM diffractometer\cite{Hannon} at the ISIS facility of the Rutherford Appleton Laboratory. The sample was contained in a thin vanadium can of diameter $8$~mm and height $40$~mm and cooled in a closed-cycle refrigerator. Data were collected at 7 and 20 K, respectively below and above the magnetic phase transition, for $8$ hours each. The raw data were reduced using Mantid\cite{Arnold} and the structural model refined from the published structure using the EXPGUI interface\cite{toby_expgui_2001} to GSAS\cite{larson_general_2004}.

Inelastic neutron scattering measurements were performed on the LET time-of-flight direct geometry spectrometer\cite{BewleyLET}, also at ISIS. The sample was contained in a thin aluminium can of diameter $15$~mm and height $45$~mm and cooled in a helium cryostat. The data were collected at a series of temperatures between $T = 5$ and 70 K, well below the structural phase transition, for approximately 2 hr each with $E_i=2.76, 5.00, 12.00$~meV using the rep-rate multiplication method\cite{Russina09,Russina10,Nakamura}. The data were reduced using the Mantid-Plot software package\cite{Arnold}. The raw data were corrected for detector efficiency and time independent background following standard procedures\cite{Windsor}.

\begin{figure}
  \centering
  \includegraphics[width=0.95\linewidth,bb=45 35 1000 670]{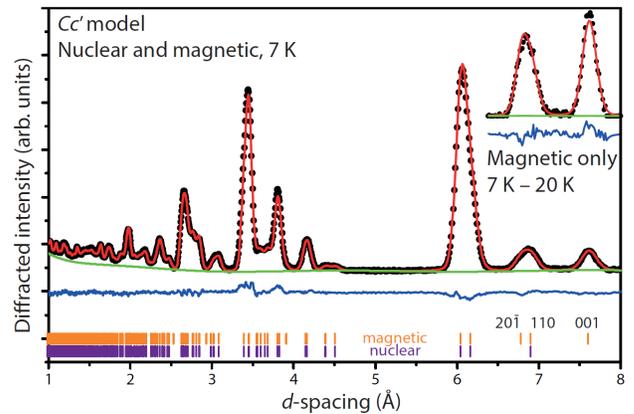}
  \caption{Powder diffraction data for \DMMnFd\ from Bank 2 ($13-21^\circ$) on GEM. The main figure shows a refinement of the nuclear and magnetic intensity, with tick marks showing the positions of the nuclear and magnetic peaks; the inset shows a refinement of the magnetic model alone to the difference between data above and below the N\'eel temperature.}
  \label{fig:diffraction}
\end{figure}

\begin{figure}
    \begin{center}
        \includegraphics[width=1\linewidth,bb=75 35 2180 2715]{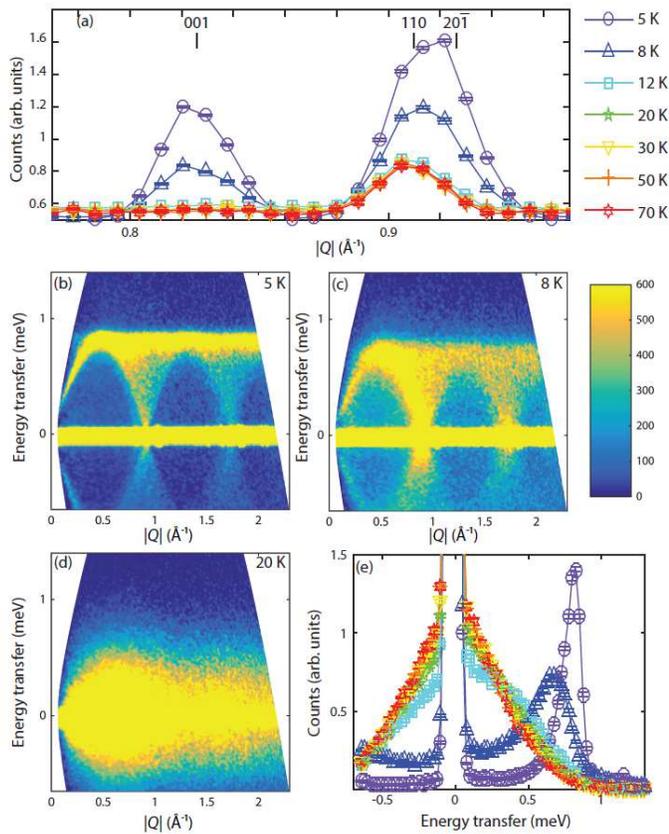}
    \end{center}
    \caption{\label{fig:multiT} Inelastic neutron scattering measured from \DMMnFd\ on LET, with an incident energy of $E_i=2.76$~meV. (a) A cut along the elastic line for $-0.25<E<0.25$~meV as a function of temperature, shows the appearance of the magnetic Bragg peaks ($001$), ($110$) and ($20\bar{1}$) below $T_\mathrm{N}$. The black tick marks are located according to the crystallographic cell at $T=7$~K. (b-d) Colour-coded inelastic neutron scattering intensity maps, energy transfer against momentum transfer $Q$, in arbitrary units, measured at $T=5,8$ and $20$~K. All three maps are on the same colour scale shown by the colour bar. (e) A cut through the magnetic scattering for $0.3<|Q|<0.7$\AA$^{-1}$ as a function of temperature using the same colour key as in (a).}
\end{figure}

\section{Results and Discussion}
In the diffraction measurements, two clear magnetic peaks were observed (Fig.~\ref{fig:diffraction}) on cooling below the N\'eel temperature. These were indexed as $(001)$ and the overlapping peaks $(110)$ and $(20\bar{1})$ with respect to the nuclear lattice. The absence of any change to the lattice suggests a magnetic propagation vector of $\Gamma=(0,0,0)$ and reduces the choice of magnetic space group to two possibilities, $Cc$ and $Cc'$. Both of these space groups are consistent with $G$-type (checkerboard) antiferromagnetism with respect to the pseudocubic, perovskite-like topology. For $Cc$, this would correspond to a model with magnetic moment in the $ac$ plane: given the strength of the $(001)$ peak, presumably with a substantial $a$ component. For $Cc'$, this would correspond to a model with magnetic moment parallel to $b$. Of the two possibilities, a $Cc'$ model fits the data substantially better ($N_\mathrm{obs}=1040$ $T=7$~K $Cc$: $R_{wp}=0.0438$, $Cc'$: $R_{wp}=0.0281$; for comparison $T=20$~K $R_{wp}=0.0261$). The magnetic moment refined to $3.86(2)\mu_\mathrm{B}$, considerably below the spin-only value of $5.92\mu_\mathrm{B}$, suggesting sizeable fluctuations or a weak ferromagnetic component to which neutron diffraction will be only weakly sensitive. The refined magnetic structure is illustrated in Fig.~\ref{fig:struc}.

The same magnetic peaks were observed at the elastic line in the spectroscopic measurements on LET (Fig.~\ref{fig:multiT}(a)).  The colour-coded inelastic neutron scattering intensity maps measured at selected temperatures are shown in Figure~\ref{fig:multiT}(b-d). At base temperature, clear spin waves emerge from the magnetic Bragg peaks at the elastic line, up to a maximum energy transfer of $0.8$~meV (Fig.~\ref{fig:multiT}(b)). No additional higher energy transfer features were observed up to a maximum of $12$~meV. For $T=8$~K, close to $T_N$, the excitations become less distinct (Fig.~\ref{fig:multiT}(c)), broadening in energy and $|Q|$, with a lowering of the maximum energy transfer reached, and a shifting of the spectral weight to lower energy transfers.  Above $T_N$ the previously sharply defined spin excitations appear to further broaden and shift down in energy before coalescing with a broad quasielastic signal that becomes more symmetric about the elastic line with increasing temperature (Fig.~\ref{fig:multiT}(e)).

In order to model the magnetic spectrum observed in the ordered phase, we calculated the spin wave dispersions, the spin-spin correlation function and the neutron scattering cross section using the SpinW program\cite{SPINW}. We used the Heisenberg magnetic Hamiltonian:
\begin{equation}
H=-\sum_{ij}J_{ij}\mathbf{S}_i\cdot\mathbf{S}_j,
\end{equation}
with three different exchange couplings for the nearest neighbour interactions, $J_1$, $J_2$ and $J_3$, between the divalent $S=\tfrac52$ Mn ions, as shown in Fig.~\ref{fig:struc}. As there is no sign of a spin gap in the spin wave spectrum, we neglected the single ion anisotropy term, which would generally open such a gap. The instrumental resolution as a function of energy transfer was estimated from the elastic line and this was included in the simulation.


\begin{figure*}
    \begin{center}
    \includegraphics[width=0.825\linewidth]{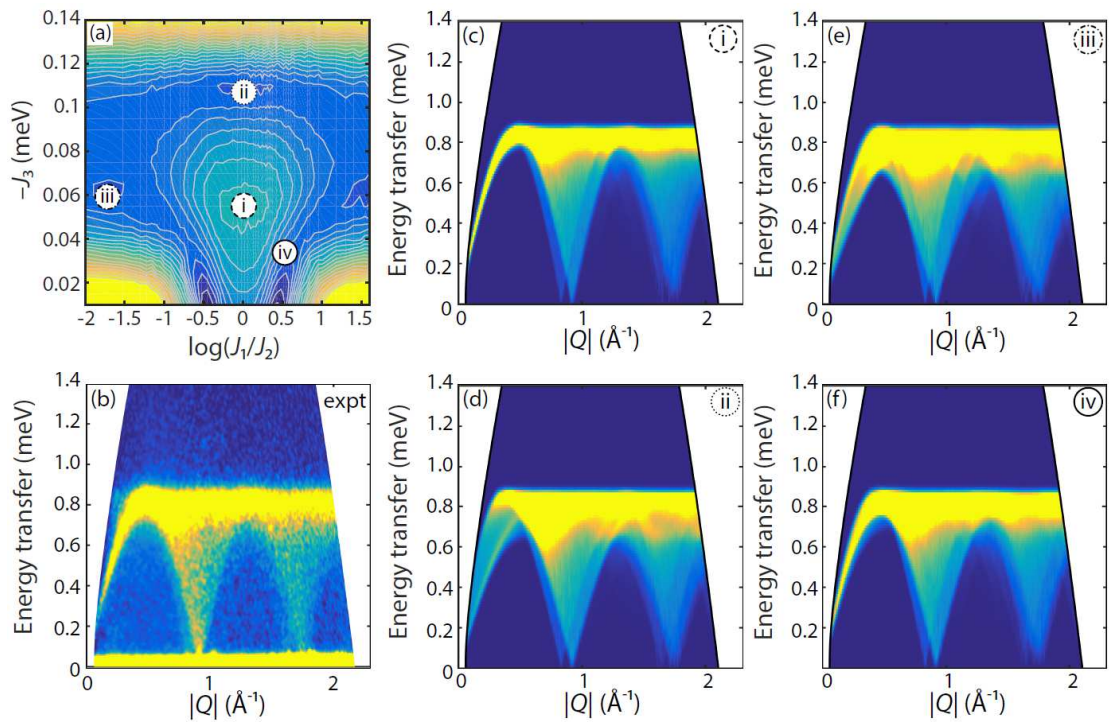}
    \end{center}
    \caption{\label{Fig:ubersims}(a) $\chi^2$-map for an energy cut at $1.2<|Q|<1.4$~\AA$^{-1}$ through the inelastic neutron scattering data from \DMMnF shown in (b) compared to spinW simulations as a function of $\log_{10}(J_1/J_2)$ vs $J_3$. (c), (d) (e) and (f) are spinW simulations for different $J_1$, $J_2$ and $J_3$ values corresponding to positions (i), (ii), (iii) and (iv) shown in the map in panel (a).}
\end{figure*}

\begin{figure}
    \begin{center}
    \includegraphics[width=0.85\linewidth]{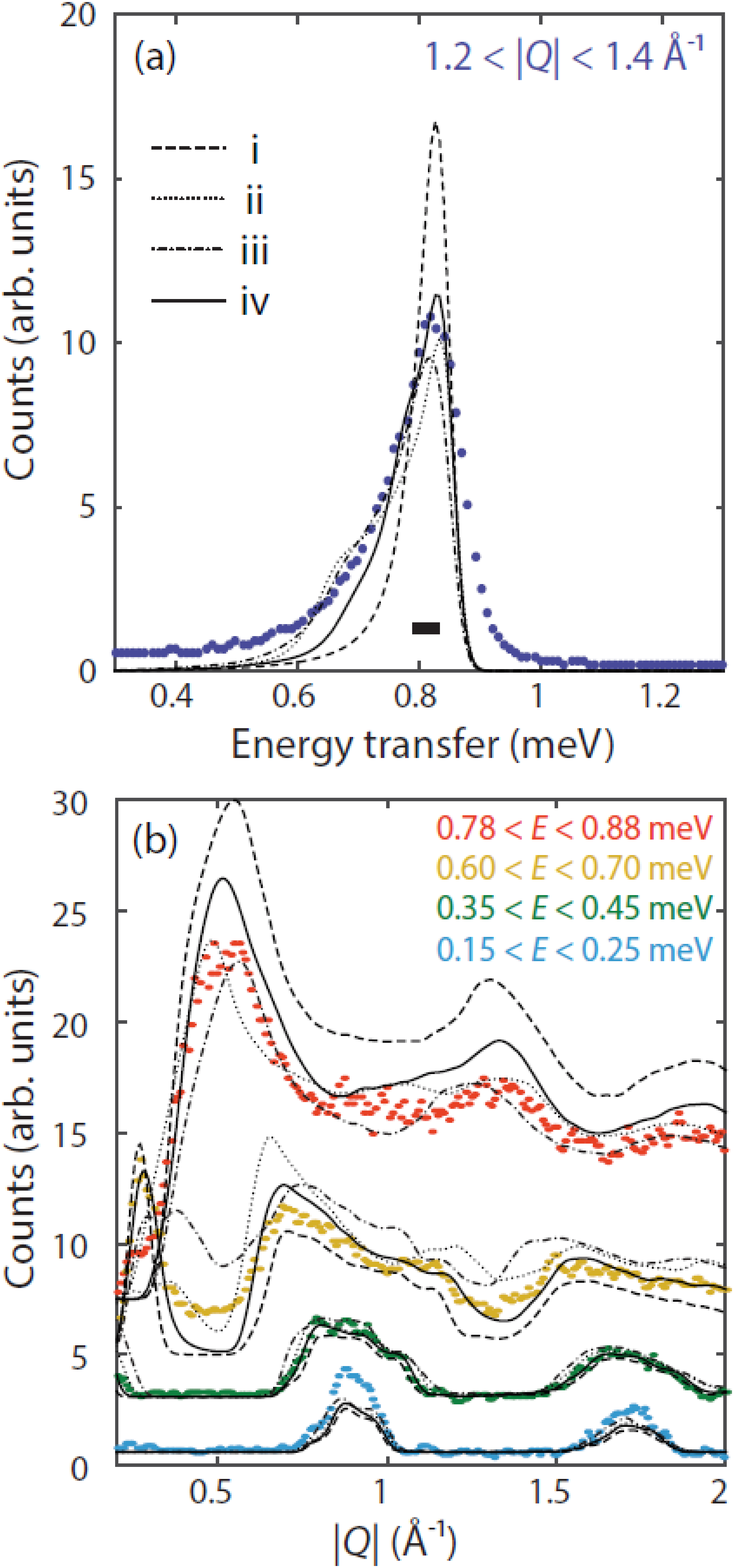}
    \end{center}
    \caption{\label{Fig:uberfits}(a) compares an energy cut through the inelastic neutron scattering data from \DMMnFd\ at $1.2<|Q|<1.4$~\AA$^{-1}$ with the same cuts through the simulations in Fig.~\ref{Fig:ubersims}(c), (d), (e) and (f); while (b) compares the different simulated results (dashed lines) to cuts along $|Q|$ for $0.78<E_T<0.88$~meV (red), $0.6<E_T<0.7$~meV (yellow), $0.35<E_T<0.45$~meV (green) and $0.15<E_T<0.25$~meV (blue). These cuts have been offset vertically for clarity. The horizontal bar in (a) represents the instrumental energy resolution.}
\end{figure}

As a starting point, given that the three nearest neighbour Mn--Mn distances are identical in the high-temperature phase and vary very little in the low-temperature phase ($d_1=5.97$~\AA, $d_2=6.18$~\AA\ and $d_3=6.21$~\AA), we took the three nearest interactions to be identical in strength, and used $J_1=J_2=J_3=-0.04$~meV, corresponding to the estimate from fits to the inverse magnetic susceptibility\cite{WangX}. However, the simulation with these values substantially underestimated both the maximum energy transfer observed in the excitations, giving only $0.58$~meV, and the width of the band maximum. Instead, a more reasonable agreement to the data was obtained for $J_1=J_2=J_3=-0.0562$~meV (Fig.~\ref{Fig:ubersims}(c)). However, while this reproduced  much of the general form of the $|Q|$ dependence of the excitations, it underestimated their energy breadth, and also the spectral weight at different energy transfers.

One of the defining features of the inelastic neutron scattering data is the energy width of the excitations observed at the van Hove-like maximum. This width is given by the energy resolution of the instrument (shown as a solid horizontal bar in Fig.~\ref{Fig:uberfits}(a)) combined with the effect of powder averaging over varying maxima in the dispersions along different directions in reciprocal space, which are dictated by the relative strengths of the three nearest neighbour interactions. Therefore, we evaluated a chi-squared map for an energy cut with $1.2<|Q|<1.4$~\AA$^{-1}$ through the different $J_1$, $J_2$, $J_3$ spinW simulations, as shown in Fig.~\ref{Fig:ubersims}(a). In order to obtain the correct energy for the band maximum, the total $J_1+J_2+J_3$ was constrained to equal $-0.17$~meV in the simulations. Given the similar geometries of the $J_1$ and $J_2$ paths it is perhaps to be expected that the plot is essentially symmetric about the $J_1=J_2$ line. Inspection of Fig.~\ref{Fig:ubersims}(a) shows that $J_1=J_2=J_3=-0.0562$~meV is actually the position of a local maximum in chi-squared, and reinforces the requirement for the interactions to be non-equal.

We subsequently used this map to inform our search for the optimal values of $J_1$, $J_2$ and $J_3$ by focussing on the positions of the local minima. The global minimum occurs for $J_3=0$, but this demonstrates why such a map, focussing on the agreement with one energy cut, can only be used as guidance. This particular fitting method was chosen as a compromise between accuracy and speed, in order to enable a search over a broad parameter space. When the full simulation is performed, and cuts along $|Q|$ are considered, it becomes clear that for $J_3=0$ the inelastic scattering emerges from the elastic line at values of $|Q|$ lower than those seen in the data (not shown here).

Panels (c), (d), (e) and (f) of Figure~\ref{Fig:ubersims} show spinW simulations for the points labelled i, ii, iii and iv in panel (a). All four reproduce the general form of the inelastic neutron scattering data presented in (b), giving a van Hove-like maximum at $\sim0.8$~meV, and spin waves emerging from the magnetic Bragg peaks at ($001$) and the overlapping ($110$) and ($20\bar{1}$), with greater spectral weight at the latter. Simpler comparisons can be made by considering one dimensional cuts: Fig.~\ref{Fig:uberfits}(a) as a function of energy transfer for $1.2<|Q|<1.4$~\AA$^{-1}$, and (b) as a function of $|Q|$ for a series of different energy transfers. For all four models, Fig.~\ref{Fig:uberfits}(a) shows that the cut-off above the van Hove-like maximum is sharper than that seen in the data (blue points), which may be related to an oversimplification in the functional form of the instrumental energy resolution used in the simulations. The energy cuts through the different model simulations then present different modulated peaks which give rise to the overall peak broadening, showing clear differences between the models. However, it is difficult to choose between these based on the powder inelastic neutron scattering data, and would instead require single crystal measurements to allow us to identify the different spin wave dispersion maxima along different directions in reciprocal space.

Turning to the cuts in Fig.~\ref{Fig:uberfits}(b), which are presented as a function of $|Q|$ for different energy transfers, the greatest variation between the different models is seen for the cut at relatively high energy transfers between $0.6$ and $0.7$~meV (yellow points). For this cut, only models i and iv, shown in Fig.~\ref{Fig:ubersims} panels (c) and (f) respectively, manage to approximately reproduce the strong sharp peak seen below $|Q|=0.3$~\AA$^{-1}$. While model iv gives a better agreement than model i to the highest energy cut. This model corresponds to $J_1=-0.1026$~meV, $J_2=-0.0324$~meV and $J_3=-0.035$~meV, and while this slightly overestimates the intensity for $0.78<E_T<0.88$~meV, and underestimates it for $0.15<E_T<0.25$~meV (something common to all four models and possibly related to the strength of the elastic scattering at the magnetic Bragg peaks which is not captured by the spinW simulations), it gives the best overall agreement with all four data cuts. Although, as discussed above, an equally good fit would be obtained by reversing the values of $J_1$ and $J_2$, this assignment is more likely, since in this model the shortest Mn--Mn distances correspond to the strongest exchange. This simple heuristic, however, is not always correct\cite{Stride}. We note further that the shortest linker has two dihedral angles of the same sign, whereas for the other two, the dihedral angles are of opposite sign. In other words, the two closest Mn atoms joined by a formate linker, corresponding to $J_1$, are on the same side of the plane of the formate ion, whereas for $J_2$ and $J_3$ the Mn atoms are on opposite sides of the plane. These geometric details are at the limit of what can robustly be resolved from the powder diffraction data. They suggest that further investigation, perhaps of single crystals, would be useful to search for geometric reasons for the differences in coupling constant.

We have investigated the possibility of spin canting, as predicted by the observation of hysteresis in magnetisation measurements\cite{WangX}. For the $J_1$, $J_2$, $J_3$ values of model iv, a canting of the spins away from the $b$ axis raises the ground state energy. It is probable that the canting is due to a weak Dzyaloshinskii-Moriya interaction, but a canting angle of $0.08^\circ$, as predicted by Wang et al.\cite{WangX}, implies an interaction strength of order $0.1$~$\mu$eV, which would be very difficult to identify using inelastic neutron scattering.

\section{Conclusions}
To summarize, in its magnetically ordered phase \DMMnFd\ is a G-type antiferromagnet with magnetic moments pointing along the crystallographic $b$ axis. The spin wave excitations have been measured using inelastic neutron scattering, and 
used to parameterize the magnetic Hamiltonian in this multiferroic metal organic framework material. Unexpectedly, the exchange constant is not the same for the six nearest neighbours: instead, the structure consists of relatively strongly coupled zig-zag chains that interact more weakly with one another, with an exchange constant ratio of $3:1$ (red:yellow/blue links in Fig.~\ref{fig:struc}). Our results suggest that orbital ordering may play an important role in the magnetic behaviour of this family of compounds. They further underline the importance of magnetic spectroscopy in elucidating the magnetic properties of novel materials such as metal-organic frameworks.

\begin{acknowledgements}
The authors thank STFC for the award of beam time at ISIS Neutron and Muon Source. A. E. P. is grateful to EPSRC for funding (EP/L024977/1), and to Dr David Palmer (CrystalMaker Software Ltd.) for assistance in preparing Figure 1. H. D. D. acknowledges QMUL, SepNet and the STFC for funding her studentship.
\end{acknowledgements}

\bibliography{dmmnf}

\end{document}